\documentclass[prb,twocolumn,amsmath,amssymb,aps,superscriptaddress]{revtex4-1}
\usepackage{amsmath}
\usepackage{graphicx}
\usepackage{dcolumn}
\usepackage{bm}
\usepackage{float}
\usepackage{url}
\usepackage{color}

\begin{document}

\title{Optical conductivity of a quantum electron gas in a Sierpinski carpet}

\author{Edo van Veen}
\email{e.vanveen@science.ru.nl}
\address{Radboud University, Institute for Molecules and Materials, NL-6525 AJ Nijmegen, The Netherlands}
\address{School of Physics and Technology, Wuhan University, Wuhan 430072, China}

\author{Andrea Tomadin}
\address{Istituto Italiano di Tecnologia, Graphene Labs, Via Morego 30, I-16163 Genova, Italy}

\author{Marco Polini}
\address{Istituto Italiano di Tecnologia, Graphene Labs, Via Morego 30, I-16163 Genova, Italy}

\author{Mikhail I. Katsnelson}
\address{Radboud University, Institute for Molecules and Materials, NL-6525 AJ Nijmegen, The Netherlands}

\author{Shengjun Yuan}
\email{s.yuan@whu.edu.cn}
\address{School of Physics and Technology, Wuhan University, Wuhan 430072, China}
\address{Radboud University, Institute for Molecules and Materials, NL-6525 AJ Nijmegen, The Netherlands}

\date{\today}

\begin{abstract}
Recent advances in nanofabrication methods have made it possible to create complex two-dimensional artificial structures, such as fractals, where electrons can be confined. The optoelectronic and plasmonic properties of these exotic quantum electron systems are largely unexplored.
In this article, we calculate the optical conductivity of a two-dimensional electron gas in a Sierpinski carpet (SC). The SC is a paradigmatic fractal that can be fabricated in a planar solid-state matrix by means of an iterative procedure.
We show that the optical conductivity as a function of frequency (i.e.~the optical spectrum) converges, at finite temperature, as a function of the fractal iteration.
The calculated optical spectrum features sharp peaks at frequencies determined by the smallest geometric details at a given fractal iteration.
Each peak is due to excitations within sets of electronic state-pairs, whose wave functions are characterized by quantum confinement in the SC at specific length scales, related to the frequency of the peak.
\end{abstract}

\pacs{PACS}

\maketitle

\section{Introduction}

Recent advances in nanofabrication methods have made it possible to create multi-scale two-dimensional (2D) structures, which are geometrically defined down to the nanometer scale, and yet feature excellent electronic quantum conduction properties on micrometer length scales.
For example, artificial lattices can be fabricated to study many phenomena in a highly tunable environment.~\cite{polini2013artificial}
More generally, nanolithography methods can yield high-quality 2D semiconductor heterostructures with a spatial resolution on the order of ten nanometers.~\cite{PhysRevB.79.241406}
Moreover, bottom-up nanofabrication methods such as nanocrystal self-assembly have been used to make self-similar structures.~\cite{shang2015assembling, C6CC04879J} These multi-scale quantum systems are naturally expected to host a wealth of unexplored optoelectronic phenomena, originating from the interplay between electronic states extending over the whole structure and states localized in the vicinity of the smallest details of the geometry.

Self-similar geometric fractals are a well-known family of multi-scale systems which can be obtained by simple iterative procedures~\cite{mandelbrot1983fractal} and are thus well suited to theoretical and experimental investigations.
Mathematically, the most striking characteristic of a fractal is that a non-integer dimension, called the Hausdorff dimension, can be defined in the limit of infinite iterations.
From an optoelectronic perspective, instead, the theory is challenging because these systems are extended and cannot be easily treated as single emitters coupled to radiation, yet they are not periodic, so that a classification of electronic states based on the Bloch theorem is not possible either.
Some analytical solutions to the Schr{\"o}dinger equation for finitely ramified fractals have been found.~\cite{PhysRevB.28.3110}
Moreover, there has been substantial work on random walks~\cite{rammal1983random}, transport~\cite{garcia2017self, PhysRevB.60.13444} and weak antilocalization~\cite{PhysRevB.94.161115} in fractals.
These papers generally focus on finding signatures of fractality in measurable physical properties. For example, quantum transport calculations unveiled a relation between the Hausdorff dimension of a planar fractal and its conductance fluctuations.~\cite{PhysRevB.93.115428}
Self-similar antennas have been designed~\cite{hohlfeld_fractals_1999}, extending the concept of the well-known log-periodic antennas, but, to the best of our knowledge, no theoretical study has ever addressed the electromagnetic properties of quantum electron systems in a fractal structure.

In this article, we calculate the optical conductivity of a two-dimensional quantum electron gas (2DEG) roaming on a Sierpinski carpet (SC), which is a paradigmatic planar fractal geometry easily generated by an iterative procedure.

In the first section, we discuss the model and methods used for our calculations. In the second section we show that the optical conductivity as a function of frequency (i.e.~the optical spectrum) converges to a definite profile as the fractal iteration increases, and we investigate the converged optical spectrum for different model parameters, highlighting the unexpected appearance of sharp peaks. Finally, we explain the
origin of these peaks by analyzing the contribution to the optical conductivity of sets of specific electronic state-pairs in SCs of reduced size, which are amenable to exact diagonalization.

\section{Model Hamiltonian and calculation method}

\begin{figure}
\includegraphics[width=\columnwidth]{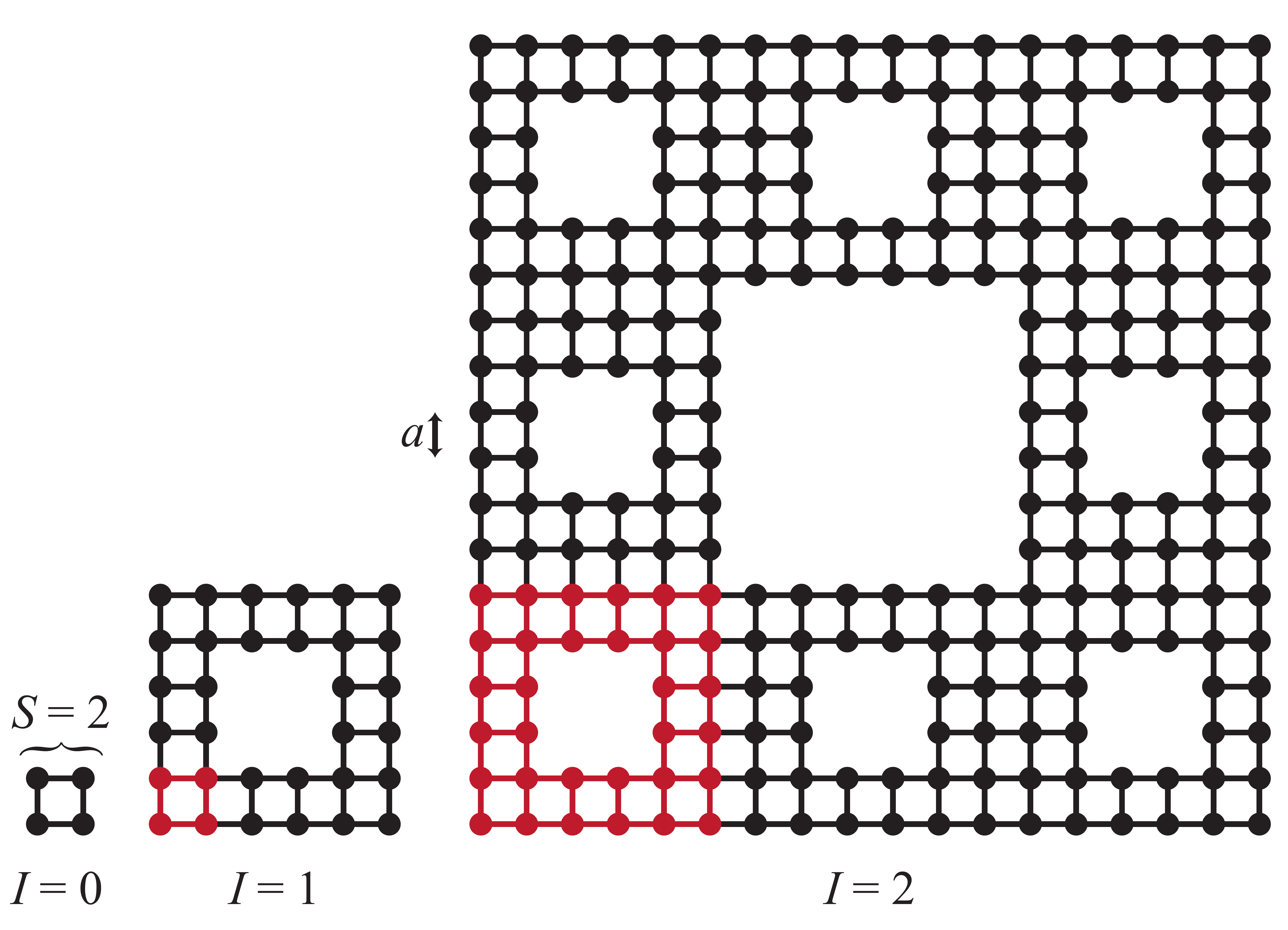}
\caption{\label{fig:sample}
The first two fractal iteration of a Sierpinski carpet on a lattice, starting with an $S \times S$ initial square with lattice constant $a$.
At each iteration $I > 0$, the unit $I-1$ is replicated $\mathcal{N}=8$ times to obtain a new structure (black) that is $\mathcal{L}=3$ times wider than the previous iteration (red). }
\end{figure}

We model a 2DEG in a SC using a single-orbital tight-binding Hamiltonian of the form
\begin{equation}\label{eq:themodel}
H = - t \sum_{\langle i,j \rangle} c^{\dagger}_i c^{\phantom{\dagger}}_j~,
\end{equation}
where $\langle i,j \rangle$ denote nearest-neighbor sites. As described in Fig.~\ref{fig:sample}, the SC is fully characterized by its starting square $S$, fractal iteration $I$, lattice constant $a$, and hopping parameter $t$. Its unit cell area is $A = a^2$, its total width is $W = S \times 3^I$ and its Hausdorff dimension is $d_H = \log_\mathcal{L} \mathcal{N} \approx 1.89$. The spectrum of the Hamiltonian is symmetric around the energy $E = 0$ and extends from $E = -4t$ to $E = 4t$.~\cite{PhysRevB.93.115428}
In our calculations, we fix the chemical potential $\mu$ in the middle of the spectrum, i.e.~$\mu = 0$, with the goal of respecting the intrinsic particle-hole symmetry of the Hamiltonian.

To compute the optical spectrum of the Hamiltonian~(\ref{eq:themodel}) on a SC, we use the tight-binding propagation method (TBPM).~\cite{PhysRevB.82.115448}
The TBPM is very efficient for large quantum systems without translational invariance, such as fractals, because it performs calculations in real space and does not require exact diagonalization.

We now briefly summarize the main steps of a TBPM calculation.
Using Kubo formula, the real part of the optical conductivity matrix $\sigma_{\alpha, \beta}$, where $\alpha$, $\beta$ are indices in real space, reads:
\begin{align}
\text{Re} \sigma_{\alpha \beta}(\omega)
= & \lim_{\epsilon \rightarrow 0^+} \frac{e^{-\hbar \omega / k_B T}}{\hbar \omega A}
\int_{0}^{\infty} e^{-\epsilon \tau} \sin \omega \tau \nonumber \\
& \times 2\text{Im} \langle \psi_2(\tau) | j_{\alpha} | \psi_1(\tau) \rangle_\beta d\tau~.
\end{align}
Here, we use the wave functions
\begin{align}
| \psi_1 (\tau) \rangle_{\beta} = & e^{-iH\tau} [1 - f(H)] j_{\beta} | \psi(0) \rangle, \\
| \psi_2 (\tau) \rangle = & e^{-iH\tau} f(H) | \psi(0) \rangle~,
\end{align}
with the Fermi-Dirac distribution operator
\begin{equation}
f(H) = \frac{1}{e^{\beta(H-\mu)} + 1}~,
\end{equation}
(with $\mu=0$), and the current operator
\begin{equation}
j_{\alpha} = - \frac{ie}{\hbar} \sum_{i,j} t (\mathbf{r}_j -
\mathbf{r}_i)_{\alpha} c^{\dagger}_i c^{\phantom{\dagger}}_j~.
\end{equation}
In this method, $| \psi(0) \rangle$ is an initial random state
\begin{equation}
|\psi(0)\rangle = \sum_i a_i \left| i \right\rangle~,
\end{equation}
where $a_i$ are random complex numbers with $\sum_i |a_i|^2 = 1$.
The time evolution $e^{-iH\tau} | \psi \rangle$  is computed numerically using Chebyshev polynomial decomposition. For a reliable result, we take the average over multiple initial random states. Because of the symmetries of the system, $\sigma = \sigma_{xx} = \sigma_{yy}$.
We use units of $\sigma_0 = e^2 / (4 \hbar)$.

The density of states (DOS) $\rho(E)$ can be calculated with TBPM as well, using the formula \cite{PhysRevE.62.4365, PhysRevB.82.115448}
\begin{equation}
\rho(E) = \frac{1}{2\pi} \int_{-\infty}^{\infty} e^{iE\tau} \langle \psi(0) |e^{-iH\tau}| \psi(0) \rangle d\tau.
\end{equation}
\begin{figure}
\includegraphics[width=\columnwidth]{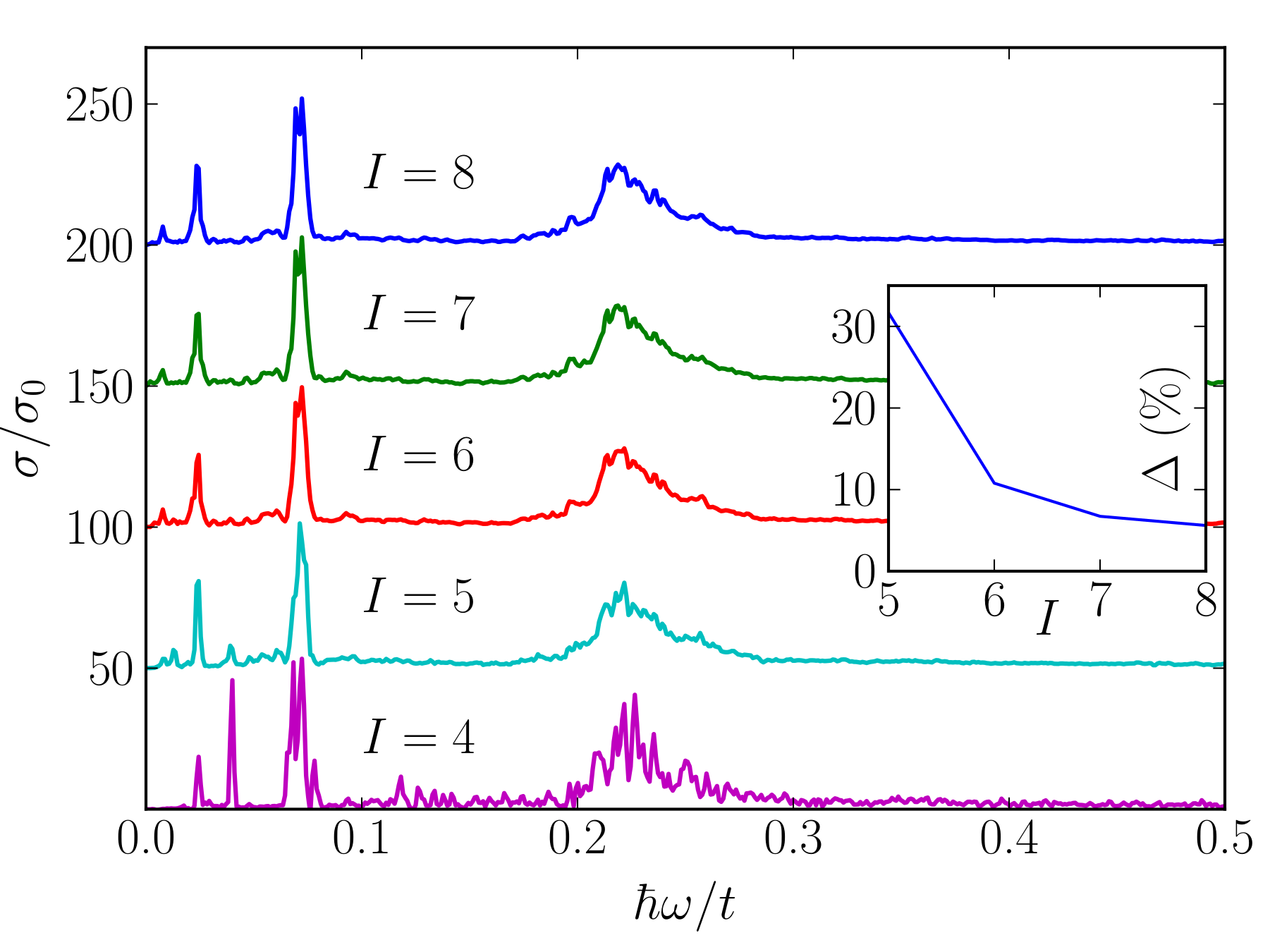}
\caption{\label{fig:compare_ac}
Optical spectrum at $S=1$ and increasing fractal iteration $I$.
(Graphs are progressively offset by $50\sigma_0$ for clarity.)
The three highest peaks for $I=5$ are already very close to the converged result for $I=7$.
The inset shows the relative difference between the conductivities at subsequent iterations, $\Delta(I) = \int |\sigma^{(I)}(\omega) - \sigma^{(I-1)}(\omega)| d\omega \ /  \int \sigma^{(I)}(\omega) d\omega$.
This quantity decreases with $I$ to $\Delta(I) \lesssim 5\%$ for $I \ge 8$. We expect a residual nonzero difference partly due to the fact that the limit of a perfect fractal has not been reached yet, and partly because we are using a finite number of random states for the TBPM calculations, resulting in some statistical fluctuations.
}
\end{figure}
\begin{figure}
\includegraphics[width=\columnwidth]{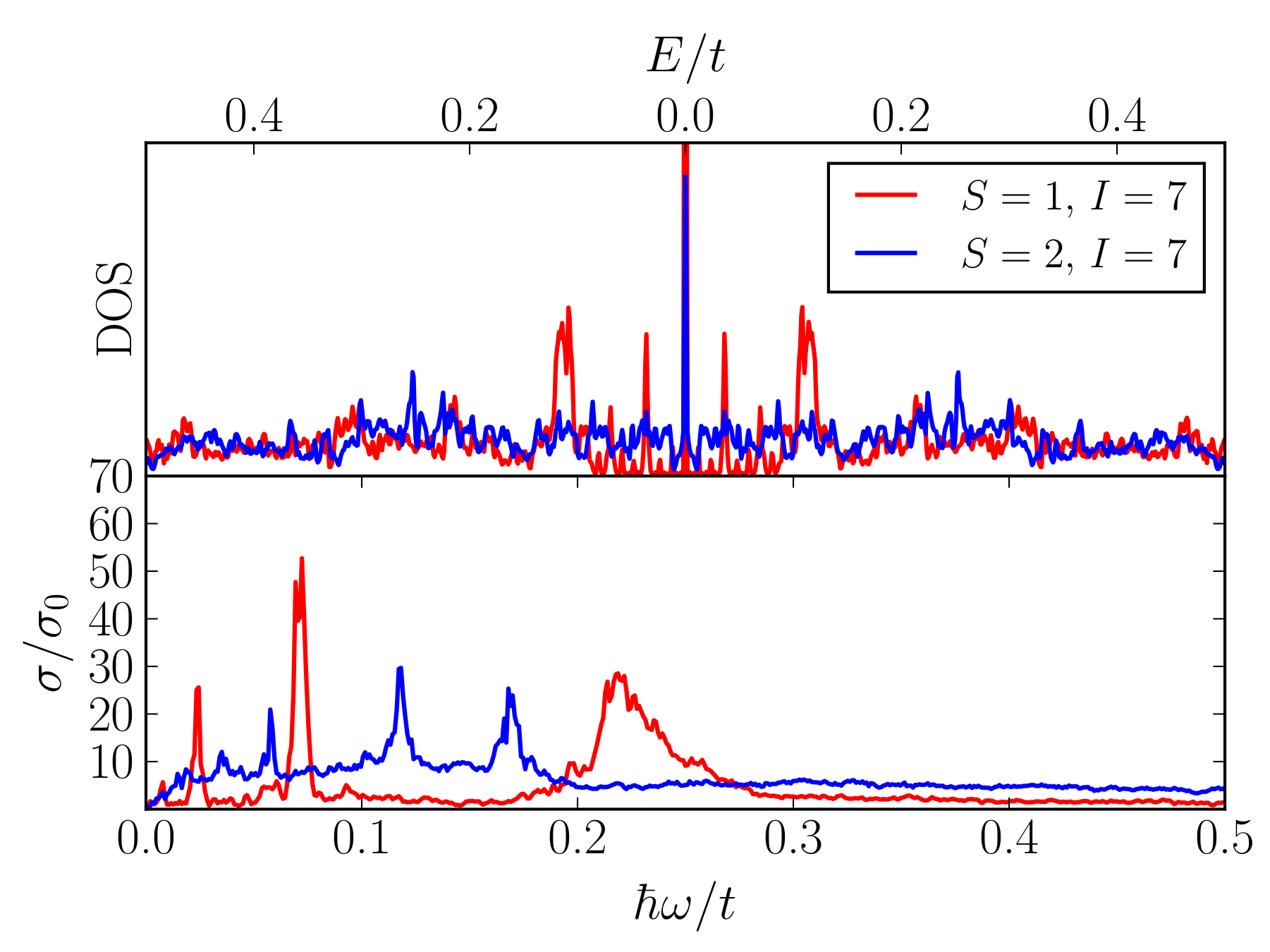}
\caption{\label{fig:dos_ac}
Converged DOS (top) and optical spectrum (bottom) for SCs with $S=1$ (red), $S=2$ (blue), and $I = 7$. }
\end{figure}
\section{Convergence and parameter dependence of the optical spectrum}

In Fig.~\ref{fig:compare_ac} we show the optical spectrum at different fractal iterations $I$.
It is remarkable that, as the total width $W$ of the SC increases, the optical spectrum maintains its overall profile.
Indeed, by comparing the results at $I=7$ and $I=8$, we conclude that, for any practical purpose, the optical spectrum has converged by iteration $I=7$.

Focusing on $I = 7$, we present in Fig.~\ref{fig:dos_ac} the optical spectrum and the DOS for different sizes $S$ of the initial $I = 0$ square.
Both quantities are markedly different for $S = 1$ and $S = 2$.
Interestingly, this shows that the finest geometric structures of the SC play a substantial role in its optical response, even in the limit of very large carpets, when such structures are negligible in size.
For both investigated values of $S$, the optical spectrum is characterized by sharp peaks at low frequencies $\hbar \omega \lesssim t$.

\begin{figure}[h]
\includegraphics[width=\columnwidth]{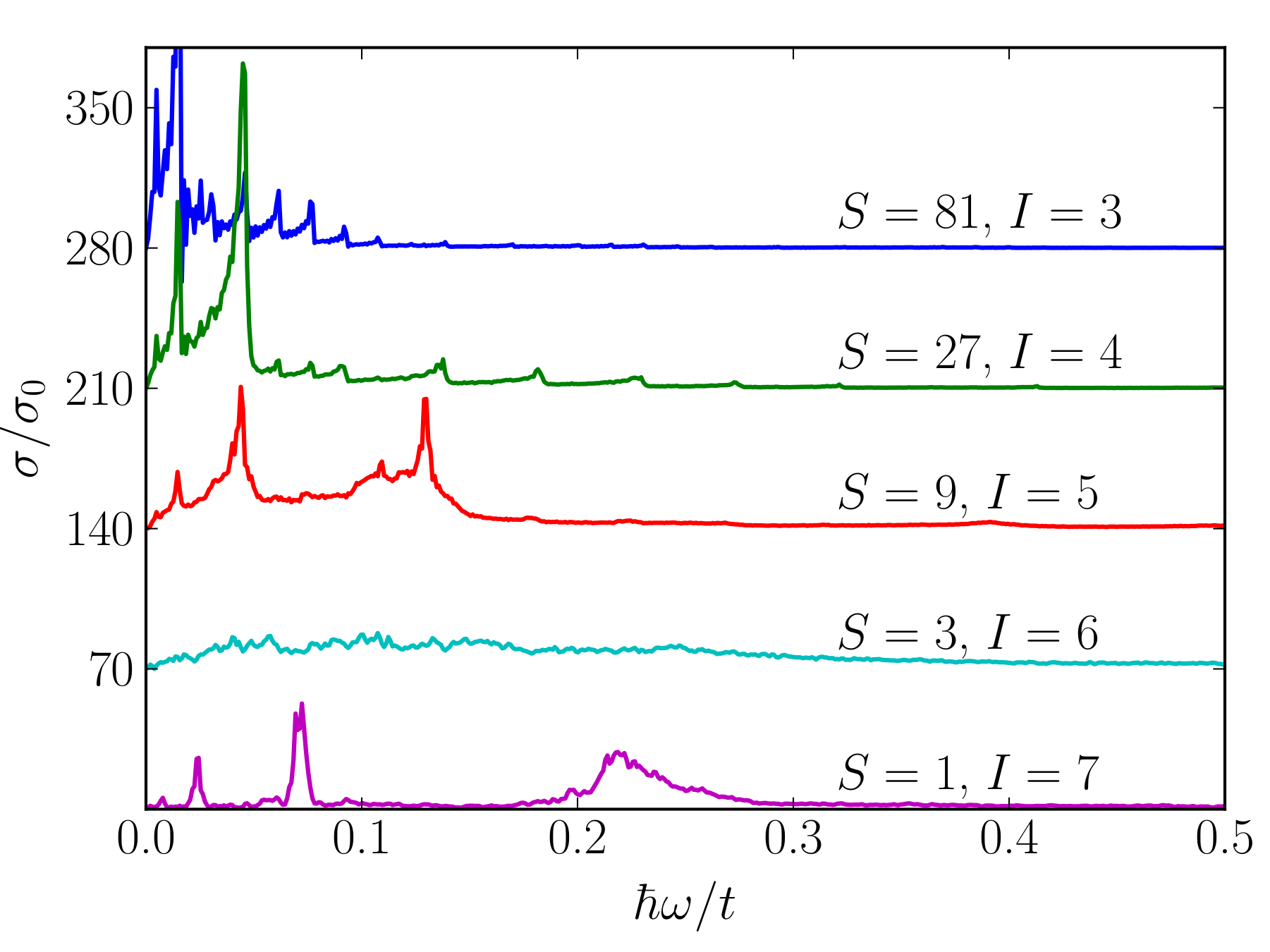}
\caption{\label{fig:etch}
Optical spectrum for fixed SC width $W = 2187$ and different fractal iteration $I$.
(Graphs are progressively offset by $70\sigma_0$ for clarity. )
To keep the width $W$ fixed, the size $S$ of the $I = 0$ square decreases as $I$ increases.
Finer geometric structure generated at higher $I$ generally introduces higher frequency peaks. }
\end{figure}

In Fig.~\ref{fig:etch} we show the optical spectrum at different fractal iterations $I$, keeping fixed the sample size $W$ and decreasing the size $S$ of the $I = 0$ square consequently.
This algorithm to the generation of the SC, known as top-down or ``intrusion,'' differs from the bottom-up or ``extrusion'' algorithm described in Fig.~\ref{fig:sample}, but  the final geometric object obtained in the limit $I \gg 1$ is the same.
This different approach represents more faithfully a physical fabrication process based on etching more and more details into a solid-state sample.~\cite{polini2013artificial}
From Fig.~\ref{fig:etch} it is apparent that increasing the detail in the sample leads to higher-frequency peaks in the optical spectrum.

\section{Origin of the peaks in the optical spectrum}

While the TBPM method allows us to calculate the optical spectrum and the DOS of SCs up to fractal iteration $I = 8$, smaller systems up to $I = 5$ are amenable to exact diagonalization.
Although the optical spectrum is not converged for $I = 5$, it already features well-defined low-frequency sharp peaks (at $\hbar \omega \simeq 0.023t$, $0.071t$, and $0.22t$) that do not shift appreciably as $I$ is increased further.
For this reason, we reckon that exact diagonalization of the SC at fractal iteration $I = 5$ can give us reliable information on the origin of the spectral peaks.

We first show that the spectral peaks cannot be understood as van-Hove-like singularities, i.e.~an enhancement of the optical response at those frequencies matching a very large set of electronic transitions.
To do so, we calculate the joint density of states (JDOS), which is given by:
\begin{equation}\label{eq:response_jdos}
\chi_{\rm JDOS} (\omega) = \frac{1}{\hbar} \sum_{nm} \frac{P_m - P_n}{\hbar\omega + E_m - E_n + i\eta}~,
\end{equation}
where, at zero temperature, $P_n = 1$ for states below the Fermi level and $P_n = 0$ otherwise. Using the JDOS, we can calculate an effective conductivity-like function
\begin{equation}\label{eq:sigmajdos}
\text{Re} \sigma_{\rm JDOS}(\omega) = - \frac{1}{\omega} \text{Im}
\chi_{\rm JDOS}(\omega)~,
\end{equation}
which quantifies the density of available electronic transitions with energy $\hbar \omega$ between state-pairs.

\begin{figure}
\includegraphics[width=\columnwidth]{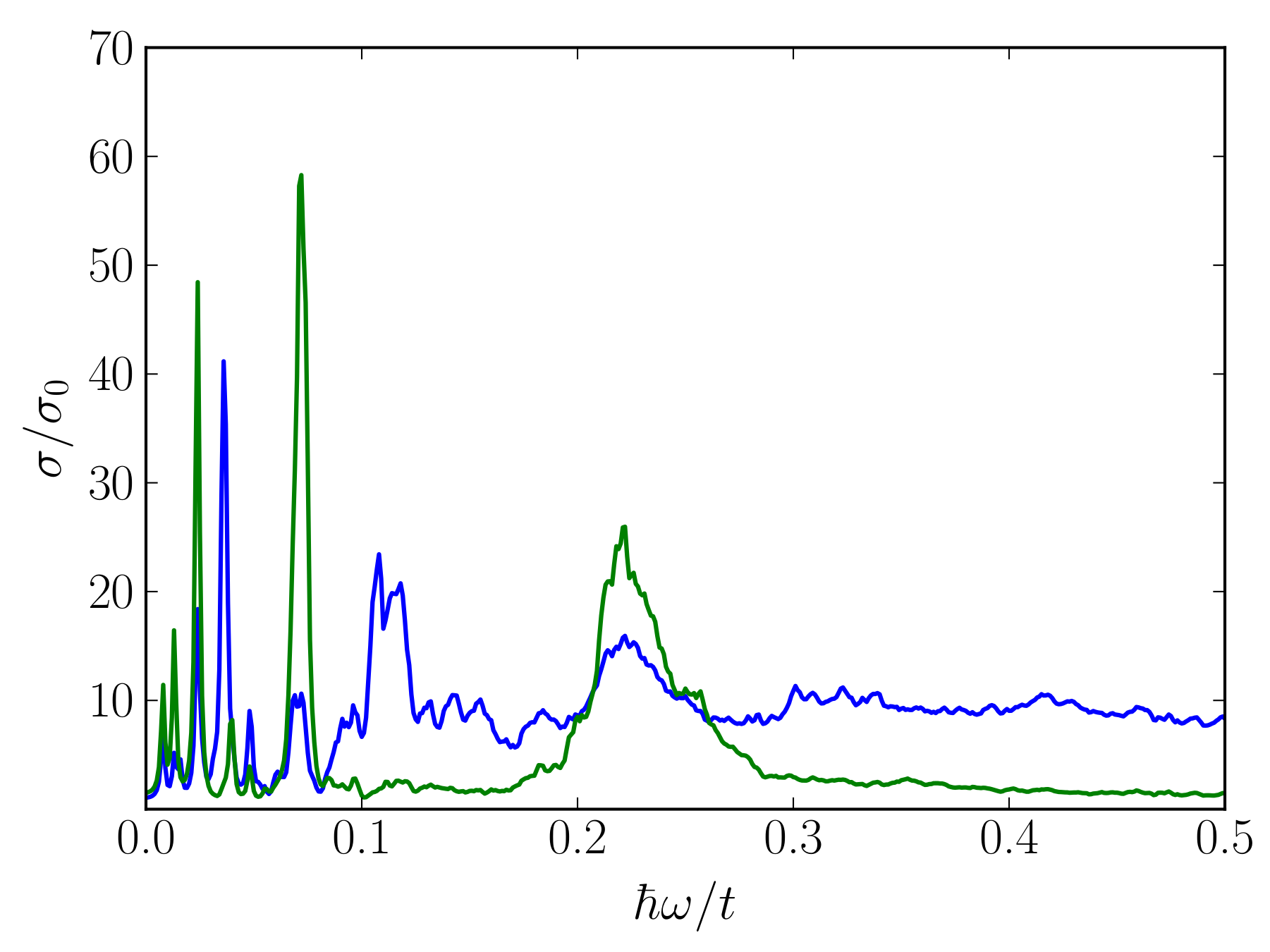}
\caption{\label{fig:jdos} Comparison of the optical spectrum (green) and the conductivity-like JDOS defined in Eq.~\ref{eq:sigmajdos} (blue) for $S = 1$ and $I = 5$.}
\end{figure}

We compare the optical spectrum and the conductivity-like JDOS extracted from Eq.~(\ref{eq:sigmajdos}) in Fig.~\ref{fig:jdos} in the specific case $S = 1$ and $I = 5$.
We clearly see that there is no substantial correlation between these two functions.
The contributions of excitations between the two peaks in the DOS at $E=-0.11t$ and $E=0.11t$ (see Fig.~\ref{fig:dos_ac}) could be expected to account for the optical conductivity peak at $\hbar \omega=0.22t$, but these contributions are washed out by those of state-pairs in which one state is around $E=0$.

\begin{figure}
\includegraphics[width=\columnwidth]{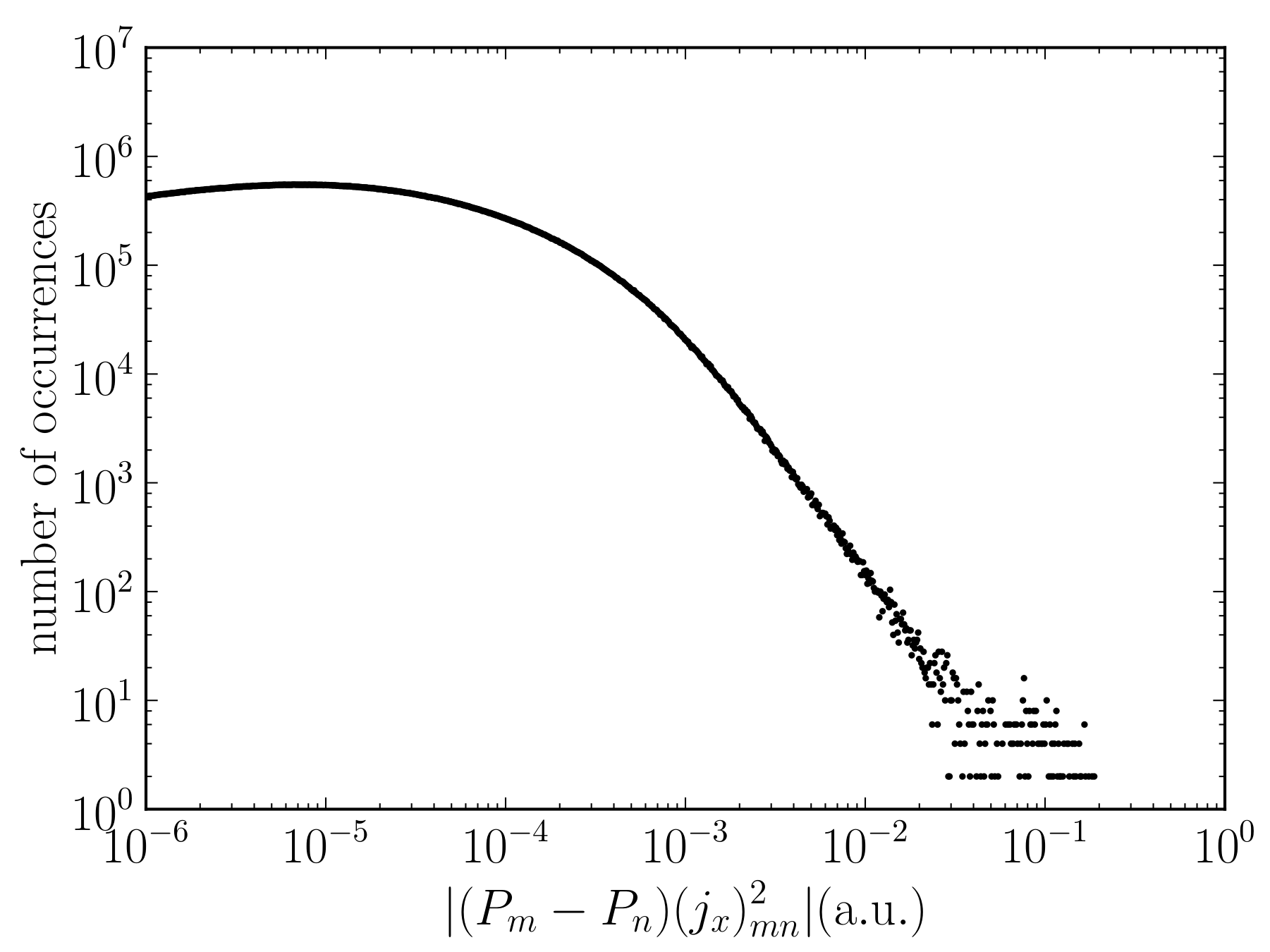}
\caption{\label{fig:jmn}
Number of occurrences of the quantity $|(P_m - P_n)(j_{x})_{mn}^2|$, using logarithmically distributed bins, calculated from the electronic spectrum in a SC with $S=1$ and $I=5$.}
\end{figure}

We now show that the spectral peaks are also not due to few, particularly effective, electronic transitions between single state-pairs. To this end, we write the current-current response function~\cite{giuliani2005quantum} in the form
\begin{equation}
\label{eq:response}
\chi_{j_{\alpha} j_{\beta}} (\omega) = \sum_{nm} Q_{mn}(\omega)~,
\end{equation}
where
\begin{equation}
Q_{mn}(\omega) = \frac{1}{\hbar A} \frac{P_m - P_n}{\hbar\omega + E_m - E_n + i\eta} (j_{\alpha})_{mn} (j_{\beta})_{nm}
\end{equation}
and $(j_{\alpha})_{mn}$ are the matrix elements of the current operator
\begin{equation}
(j_{\alpha})_{mn} = \langle \psi_m | j_{\alpha} | \psi_n \rangle~.
\end{equation}
Here, $| \psi_m \rangle$ are the eigenstates of the Hamiltonian~(\ref{eq:themodel}). 
For each matrix element, $(j_{\alpha})_{mn}$, we calculate the quantity $|(P_m - P_n)(j_{x})_{mn}^2|$, which is a measure of the strength of an electronic transition, independent of the frequency of the field which drives the transition itself.
Fig.~\ref{fig:jmn} shows the distribution of the magnitude of this quantity.
If the peaks in the optical spectrum were due to a few electronic transitions, the distribution should have a few large values with a small number of occurrences -- which is clearly not the case.

\begin{figure}
\includegraphics[width=0.7\columnwidth]{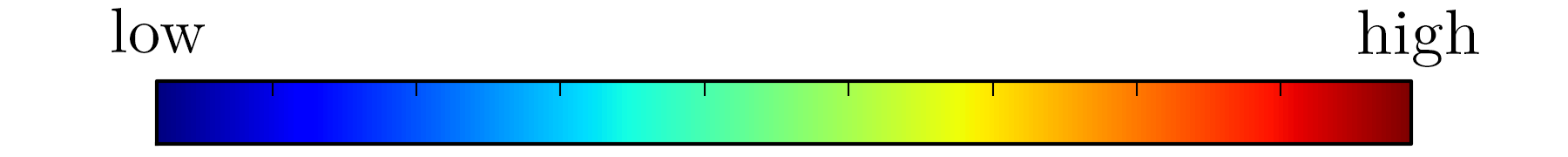}\\
\includegraphics[width=\columnwidth]{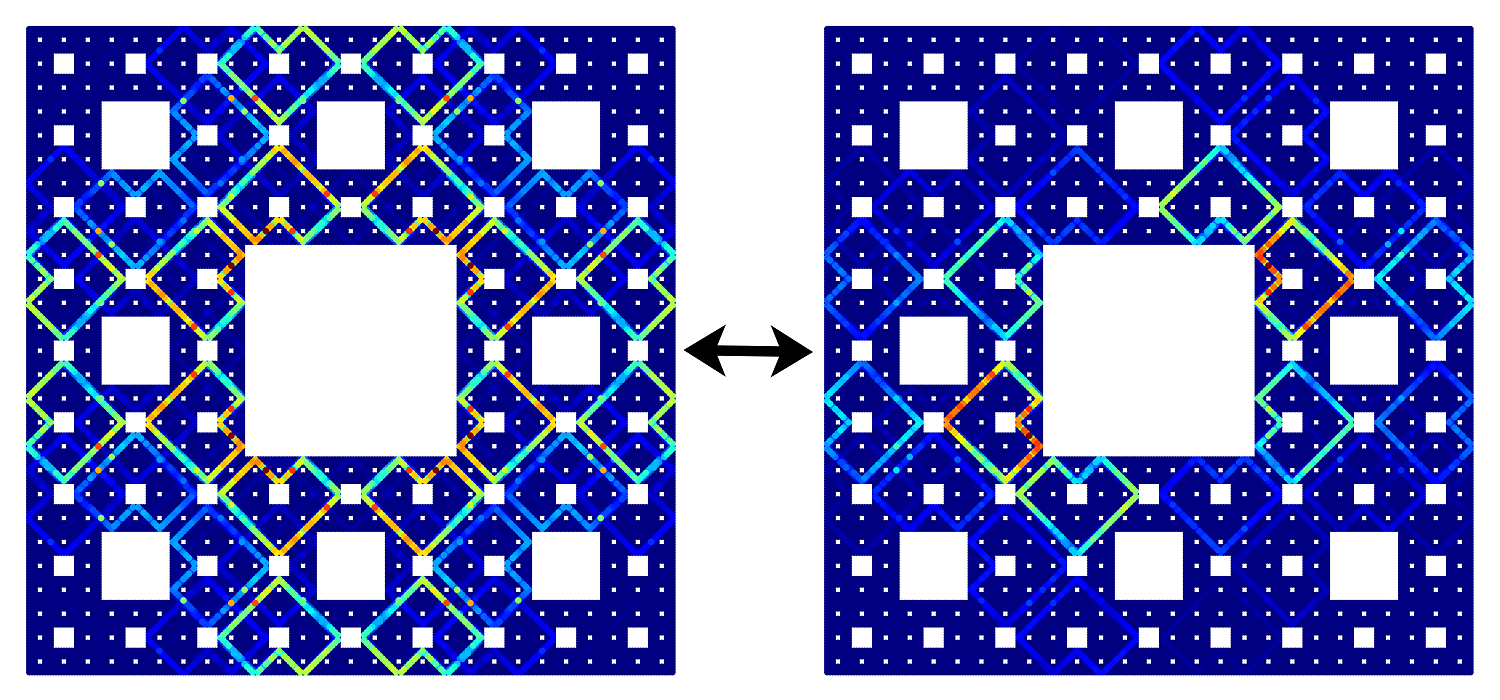}\\
\includegraphics[width=\columnwidth]{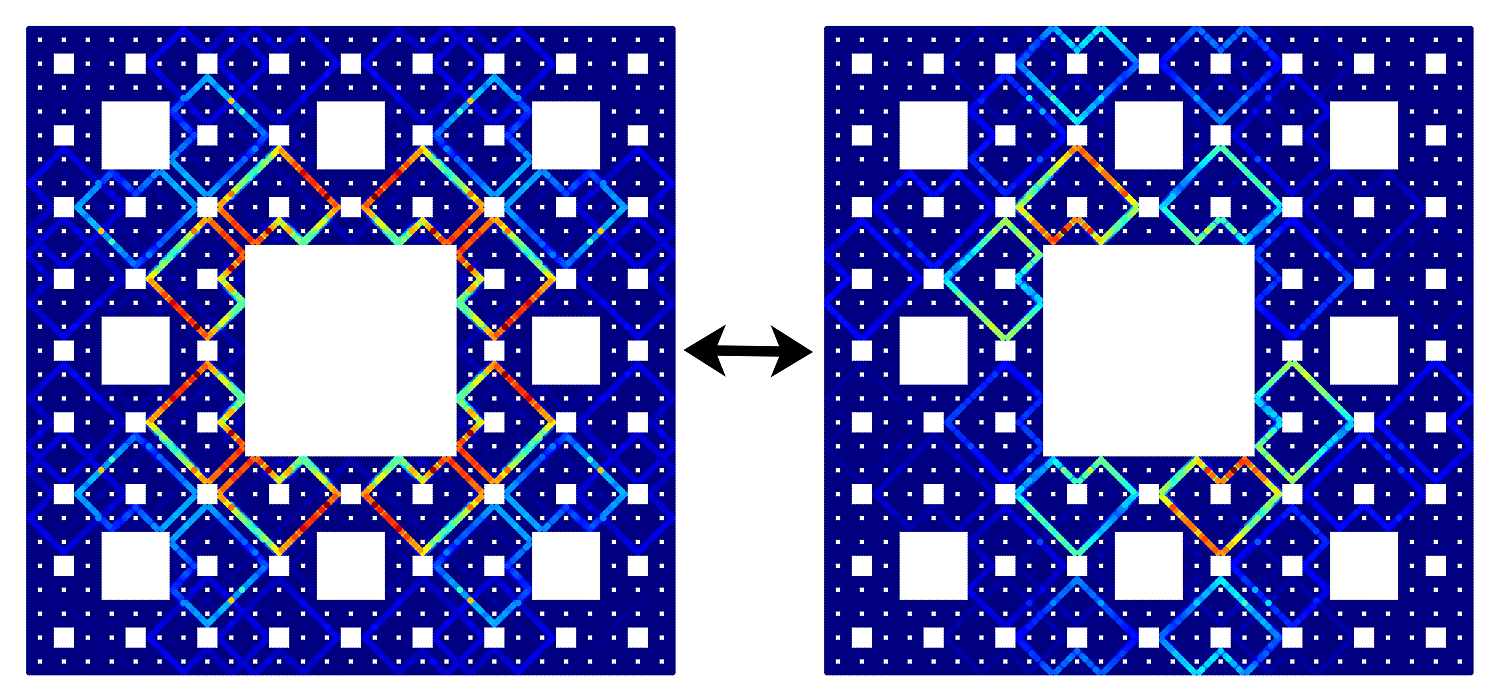}
\caption{\label{fig:wfpairs}
Two sets of top-contributing state-pairs for the peak at $\hbar \omega = 0.071t$, in a SC with $S=1$ and $I=5$. }
\end{figure}

\begin{figure}
\vspace{0.5cm}
\includegraphics[width=0.7\columnwidth]{colorbar.png}\\
\includegraphics[width=\columnwidth]{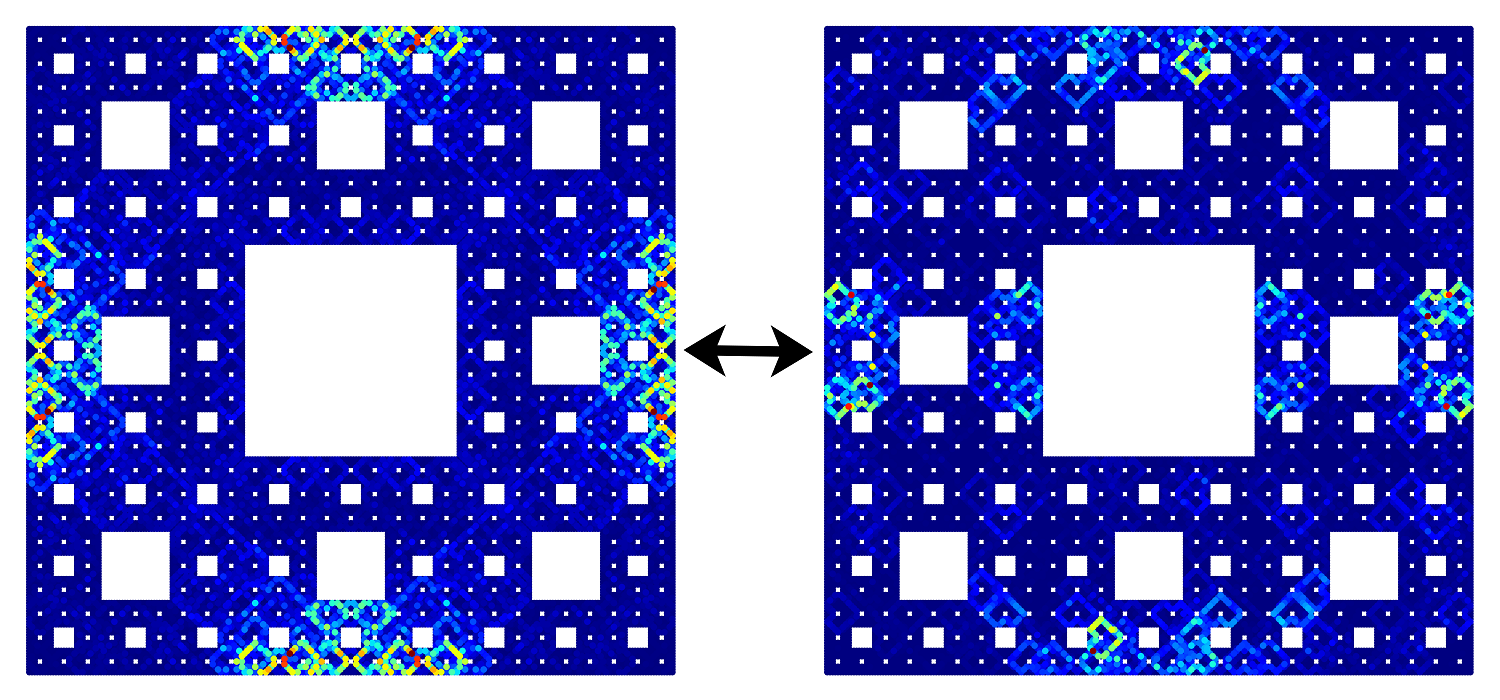}\\
\includegraphics[width=\columnwidth]{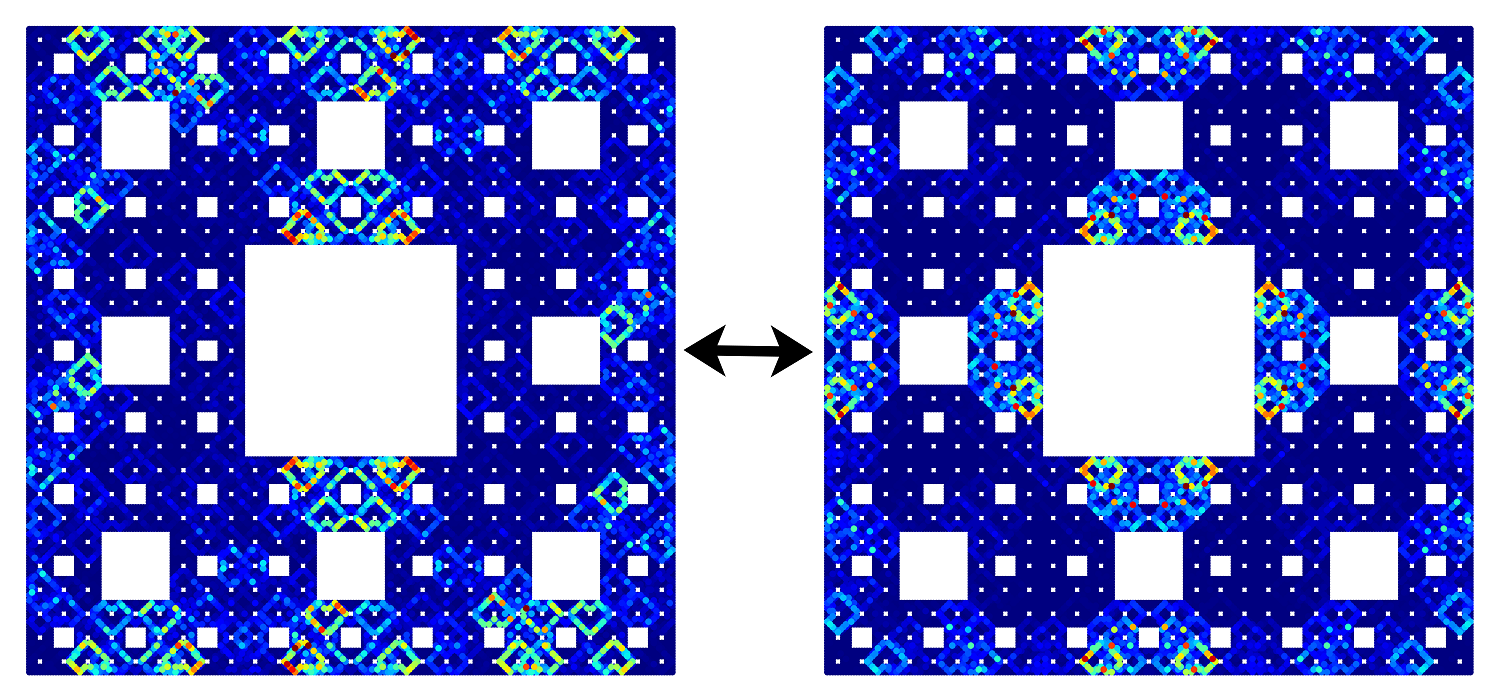}
\caption{\label{fig:wfpairs2}
Two sets of top-contributing state-pairs for the peak at $\hbar \omega = 0.220t$, in a SC with $S=1$ and $I=5$. }
\vspace{0.5cm}
\end{figure}

Summarizing the analysis above, we have ruled out that sharp peaks in the optical spectrum arise from dense, energy-localized sets of transitions, or from sparse, isolated transitions between state-pairs.
We are then left with the option that the origin of the spectral peaks are transitions between large and non-trivial sets of state-pairs, uncorrelated with the JDOS.
In the following, we characterize these sets, by directly looking at the probability density of the wave functions on the SC.
For example, the large peak in ${\rm Re}\sigma(\omega)$ at $\hbar \omega=0.071t$ in Fig.~\ref{fig:jdos} for a SC with $S=1$ and $I=5$ is due to a collection of hundreds of state-pairs, two of which are shown in Fig.~\ref{fig:wfpairs}.
These state-pairs have all nearly the same contribution to that peak in the optical spectrum and display very similar ``heart-shaped'' spatial features on the scale of the geometric details introduced by the third ($I  = 3$) fractal iteration.
Similarly, in Fig.~\ref{fig:wfpairs2} we display the state-pairs contributing most to the peak at $\hbar \omega=0.22t$ in a SC with $S=1$ and $I=5$, which display similar heart-shaped profiles, but with
length scales that are $\mathcal{L} = 3$ times shorter, on the order of the second ($I = 2$) fractal iteration.
All these wave functions show very similar profiles, corresponding to confinement at a specific fractal iteration, with higher peak frequencies being related to shorter length scales within the SC.
This behavior agrees with the results shown in Fig.~\ref{fig:etch}, i.e.~``etching'' an extra iteration into the sample generally introduces higher frequency optical peaks.

To make a more quantitative connection between the peak frequencies in the optical spectrum and the characteristic ``confinement lengths'' appearing in the electronic wave functions, we calculate the sum of probability densities, weighted by their contribution to the optical conductivity, as a function of $\omega$:
\begin{equation}
S({\bm r}, \omega) = - \sum_{mn} \frac{1}{\omega} \text{Im} \left [ Q_{mn}(\omega) \right ] |\langle {\bm r} | \psi_n \rangle|^2~.
\end{equation}
The sum is restricted to states $m$ and $n$ such that their energy difference falls within the window $\hbar(\omega - \delta \omega) < |E_m - E_n| < \hbar(\omega + \delta \omega)$, with $\hbar \delta \omega = 0.01t$.
Due to particle-hole symmetry, the result is the same for taking the probability distributions $|\langle {\bm r} | \psi_n \rangle|^2$ over the index $m$.

\begin{figure}
\vspace{0.5cm}
\includegraphics[width=0.7\columnwidth]{colorbar.png}\\
\includegraphics[width=\columnwidth]{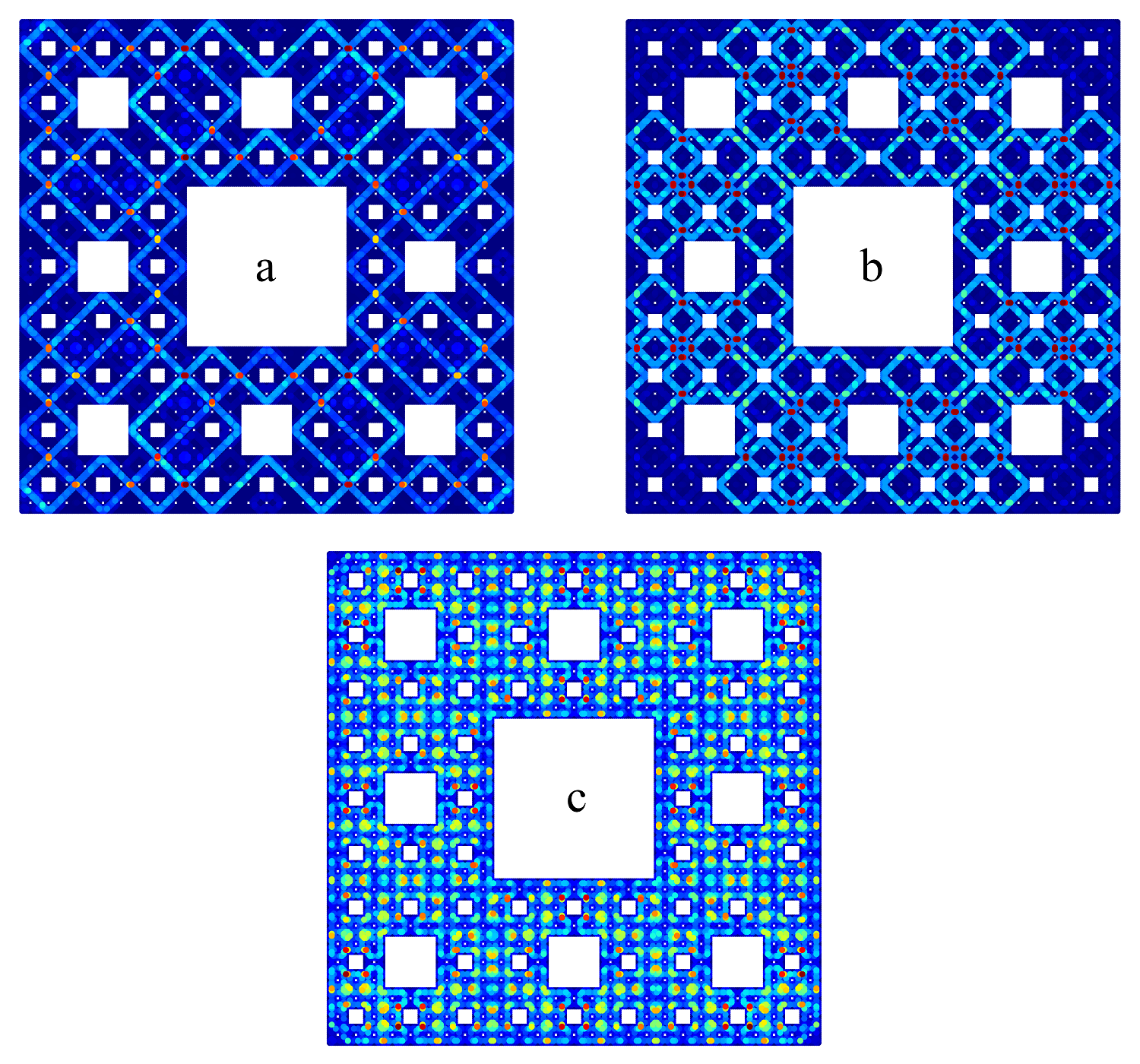}
\caption{\label{fig:weighted_sum}
Cumulative probability distributions $S({\bm r},\omega)$ of state-pairs contributing to the
peaks at (a) $\hbar \omega=0.023t$; (b) $\hbar \omega=0.071t$; and (c) $\hbar \omega=0.22t$, in a SC with $S=1$ and $I=5$. }
\end{figure}

Fig.~\ref{fig:weighted_sum} shows the spatial profile of the quantity $S({\bm r}, \omega)$ in a SC, for three values of $\omega$ corresponding to peaks in the optical spectrum.
The plots demonstrate a clear distinction in the characteristic length scale of the probability density for different frequencies.

The substantial numerical effort needed to exactly diagonalize the Hamiltonian~(\ref{eq:themodel}) on a SC hinders a more precise characterization of the state-pairs sets.
We note that the heart-shaped features of the probability density are distorted  at $\hbar \omega = 0.023t$, where the confinement length scale is on the order of the geometric details introduced by the fourth ($I = 4$) fractal iteration.
This is an artefact of the final size of the SC that we can diagonalize exactly, and we reckon that at the sixth ($I = 6$) fractal iteration the heart-shaped features would fit the SC geometry.
Moreover, for $S = 2$ a similarly thorough analysis is too expensive numerically to cover in this work.
At $I = 4$, there is already some connection between length scale and optical peak frequency, and there appears to be some extra splitting, causing two peaks per length scale.
However, the optical conductivity is not yet close enough to its converged result to make any conclusive statements.

\section{Summary}
In this work we have calculated the optical spectrum of a quantum electron gas roaming in a Sierpinski carpet.
We have shown that the optical spectrum converges to a definite profile as the fractal iteration increases.
The optical spectrum displays sharp peaks, which blue-shift as finer geometric structures are produced at higher fractal iterations.
We have pinned down the origin of these peaks to electronic transitions between set of specific state-pairs whose wave functions experience quantum confinement in the Sierpinski carpet at specific length scales.

\acknowledgements

This work was supported by the European Research Council Advanced Grant program (contract 338957) (M.I.K.) 
and by the European Union's Horizon 2020 research and innovation programme under grant agreement No. 696656 ``Graphene Flagship'' (A.T. and M.P.).
Support by the Netherlands National Computing Facilities foundation (NCF) is gratefully acknowledged.

\bibliography{bibfile}

\end{document}